\documentstyle[prd,aps]{revtex}

\begin{document}
\input psfig
\title{Perturbative evolution of conformally flat 
 initial data for a single boosted black hole}

\author{Reinaldo J. Gleiser$^1$, Gaurav Khanna $^2$,
Jorge Pullin$^3$}
\address{1. Facultad de Matem\'atica, Astronom\'{\i}a y F\'{\i}sica,
Universidad Nacional de C\'ordoba,\\ Ciudad
Universitaria, 5000 C\'ordoba, Argentina.}
\address{
2. Natural Science Division, Southampton College of Long Island
University,
Southampton NY 11968}
\address{
3. Department of Physics and Astronomy, 202 Nicholson Hall,
Louisiana State University, Baton Rouge LA 70803-4001}

\date{May 16th 2002}
\maketitle
\begin{abstract}
  The conformally flat families of initial data typically used in
  numerical relativity to represent boosted black holes are not those
  of a boosted slice of the Schwarzschild spacetime.  If such data are
  used for each black hole in a collision, the emitted radiation will
  be partially due to the ``relaxation'' of the individual holes to
  ``boosted Schwarzschild'' form.  We attempt to compute this
  radiation by treating the geometry for a single boosted conformally
  flat hole as a perturbation of a Schwarzschild black hole, which
  requires the use of second order perturbation theory. In this we
  attempt to mimic a previous calculation we did for the conformally
  flat initial data for spinning holes. We find that the boosted black
  hole case presents additional subtleties, and although one can
  evolve perturbatively and compute radiated energies, it is much less
  clear than in the spinning case how useful for the study of
  collisions are the radiation estimates for the ``spurious energy''
  in each hole. In addition to this we draw some lessons on which
  frame of reference appears as more favorable for computing black
  hole collisions in the close limit approximation.

\end{abstract}

\pacs{4.30+x}

\section{Introduction}\label{sec:intro}

There is significant interest in  obtaining
waveforms for the gravitational radiation produced in the collision of
black holes. Progress is being made on this problem both
using supercomputers \cite{grandlychallenged} and perturbative calculations
\cite{PuKy}. One of the open issues is what families of initial data 
are appropriate to represent the collision of two black holes, especially
when the latter are not far away from each other. The state of the art
of numerical simulations suggests that for some time we may not be able
to start simulations with the black holes at a sufficiently large
separation, such that one
can assume a simple linear superposition will work. This leaves open
the issue of how much ``spurious radiation'' is one introducing in the
various proposals for superpositions in the non-linear regime.
Bowen and York \cite{BoYo} (and with a different set of boundary
conditions more recently Brandt and Br\"ugmann \cite{BrBr}) studied the
problem of giving initial data for boosted and spinning holes in such
a way that a superposition is possible. They assume the spatial metric
is conformally flat and as a consequence one can superpose the
extrinsic curvatures for the two holes and still solve the momentum
constraint. One then proceeds to find a conformally flat spatial
metric for the superposed holes by solving a nonlinear elliptic
partial differential equation. The procedure achieves superposition at
the price of assuming conformal flatness of the three metrics, which
is not generically possible, and more importantly, is not possible in 
situations of interest. For instance a single spinning black hole,
described by the Kerr solution, is not known to admit conformally flat
spatial slices \cite{BoYo,GaPr}. 

A similar situation develops for the case of a single boosted black
hole. The initial data constructed by Bowen and York or Brandt and
Br\"ugmann do not correspond to those one would find on a boosted
slice of the Schwarzschild spacetime. In this paper we will refer to
these families as ``conformal boosted black hole'' Therefore if one
evolves these families of data one should find that the black hole
``settles down'' to a Schwarzschild form through the emission of
gravitational radiation.  The original purpose of this paper was to
study the emitted radiation by treating the conformal boosted hole as
a perturbation of a Schwarzschild black hole.  In a previous paper we
had carried out a similar discussion for the case of a single spinning
conformally flat hole \cite{GlNiPrPuboyo}.  We will see that the
boosted case is much more subtle than anticipated. We will be able to
evolve the spacetime, but questions will remain about the usefulness
of the results obtained, at least for the original purpose of gaining
insight into the spurious radiation content of data for black hole
collisions for interesting ranges of parameter values.

The organization of this paper is as follows. In the next section we
will discuss perturbations of a boosted black hole. We recall that
the $\ell=1$ even modes are pure gauge and therefore all the physics of
interest takes place in the $\ell=2$ modes, which can be treated
easily up to second order in perturbation theory. In section III we
will discuss the perturbative evolution and the amount of radiation
produced. We end with a discussion of the results and their implication
for the choice of frame of reference one uses in perturbative evolutions
of black hole collisions with linear momentum.

\section{Conformal  boosted black hole as a perturbation of Schwarzschild}

\subsection{Initial data in the conformal approach}

The families of initial data that we will consider in this paper are
obtained via the ``conformal approach'' to the initial value problem
in general relativity. In it, one assumes the metric to be conformally
flat $g_{ab} =\phi^4 \delta_{ab}$ and defines the conformal
extrinsic curvature $\widehat{K}_{ab} = \phi^{2} K_{ab}$. In terms of
these variables the initial value constraint equations (assuming
maximal slicing ${\rm Tr} K =0$) read,
\begin{eqnarray}
\nabla_a \widehat{K}^{ab} &=& 0\\
\nabla^2 \phi &=& -{1\over 8} {\widehat{K}^{ab} \widehat{K}_{ab} 
\over \phi^{7}}
\label{Hamil}
\end{eqnarray}
where all the derivatives are with respect to flat space.

One can construct \cite{BoYo} solutions to the first set of equations
(momentum constraint)  for a
single black hole centered at $R=0$,  with linear momentum $P_a$,
\begin{equation}\label{onehole}
\widehat{K}_{ab} = {3 \over  2 R^2} \left[ 2 P_{(a} n_{b)} -(\delta_{ab}
-n_a n_b)P^c n_c\right]\ .  \label{boyok}
\end{equation}
where $R$ is a spherical radial coordinate and $n_a$ a radial unit vector,
and both are defined in the fiducial flat space that one obtains
setting the conformal factor to unity.

Without loss of generality we may assume that $P_a$ points along the
positive z-axis, and has magnitude $P$. If we write $\widehat{K}_{ab}$
in spherical coordinates, the only non vanishing components are
\begin{eqnarray}\label{Ksphe}
\widehat{K}_{RR} &=& {3 P  \over R^2} \cos(\theta) \nonumber \\
\widehat{K}_{R\theta} &=& -{3 P  \over 2 R} \sin(\theta) = 
\widehat{K}_{\theta R}\nonumber \\
\widehat{K}_{\theta\theta} &=& -{3 P  \over 2} \cos(\theta) \nonumber \\
\widehat{K}_{\phi \phi} &=&  \sin(\theta)^2 \widehat{K}_{\theta\theta} 
\label{extriseca}
\end{eqnarray}

If we write the extrinsic curvature in terms of tensor spherical
harmonics \cite{ReWh}, we see that it consists of a pure $\ell=1$ even
term. This is reasonable, since the presence of momentum in an initial
slice configuration is determined by the presence of a ``dipole'' term
asymptotically, that makes the ADM integral
\begin{equation}
P_i = {1 \over 8 \pi } \int_\infty K_{ij} d^2 S_j
\end{equation}
nonvanishing.

From (\ref{extriseca}), we have, 
\begin{equation}\label{Kcuadrado}
\widehat{K}_{ab}\widehat{K}^{ab} = {9 P^2 \over 2 R^4} \left( 1 + 2
\cos^2 
\theta \right)
\end{equation}

We now need to solve (\ref{Hamil}), and this involves imposing
boundary conditions on $\phi$. One possibility is that given by Bowen
and York \cite{BoYo}\footnote{Alternatively, we may use a
Brill--Lindquist type boundary condition \cite{BrLi} and as a result
one obtains the ``puncture'' solutions of \cite{BrBr}.}. In this case
one chooses a certain constant $a$, solves (\ref{Hamil}) for $R
\geq a$, and imposes
\begin{equation} 
{\partial \phi \over \partial R} +{1 \over 2R} \phi =0 
\;\; \mbox{for} \;\; R= a,   \label{robin}
\end{equation}
and
\begin{equation} 
  \phi > 0 
\;\; , \;\; \lim_{R \rightarrow \infty} \phi = 1.   \label{bounds}
\end{equation}

Equation (\ref{robin}) implies that the inner boundary is an extremal
surface, as was shown in reference \cite{BoYo} and leads to a well
posed elliptic problem when used with equations
(\ref{Hamil},\ref{bounds}).  One can also use the boundary condition
(\ref{robin}) to generate initial data that obeys an isometry
condition, provided one chooses an extrinsic curvature ---different
from the one we are choosing here--- that satisfies the isometry
condition. This was explored in great detail by Cook \cite{Cook} in
his numerical work constructing solutions for the conformal factor for
multiple black holes. The symmetrization procedure yields non-trivial
results even for a single black hole as we are considering here
\cite{CoYo}. We have chosen, for simplicity, not to symmetrize the
extrinsic curvature.  Experience has shown that symmetrization does
not change significantly the amount of radiated energy in head-on
black hole collisions \cite{GlNiPrPuboost2}. Therefore our choice
should not crucially influence the central conclusions we are
attempting to obtain about the conformally flat black hole solutions.

Even for the simple form of (\ref{Kcuadrado}), in general we cannot
solve (\ref{Hamil}) exactly, and one has to resort to numerical
methods. In the present analysis, however, we will be interested in
solutions for ``small'' $P$. Since for $P=0$ the solution is
\begin{equation} 
  \phi^{(0)} = 1 + {a \over R}   
\end{equation}
we may solve (\ref{Hamil}) iteratively replacing
\begin{equation} 
 \phi = \phi^{(0)}(R) + P^2 \phi^{(2)}(R,\theta) +P^4
\phi^{(4)}(R,\theta) 
+ \cdots    
\end{equation}
in (\ref{Hamil}), and imposing  
\begin{equation} 
{\partial \phi^{(i)} \over \partial R} +{1 \over 2R} \phi^{(i)} =0 
\;\; \mbox{for} \;\; R= a, \;\; \;\;\; \lim_{R \rightarrow \infty} 
\phi^{(i)} = 0. \;\; \mbox{for} \;\; i\neq 0,   
\end{equation}

The form of (\ref{Kcuadrado}) further suggests that we expand
$\phi^{(i)}$ in Legendre polynomials $P_{\ell}(\cos\theta)$, so that
$\phi^{(2)}$ may be written as
\begin{equation} 
\label{eqphi}
 \phi = \phi^{(0)}(R) + P^2 \left[\phi^{(2)}_0(R) P_0(\cos \theta)+ 
\phi^{(2)}_2(R) P_2(\cos \theta)\right]  + {\cal{O}}(P^4).    
\end{equation}

Solving for the coefficients taking into account the boundary 
condition (\ref{robin}) we get,
\begin{eqnarray}
\phi^{(2)}_0(R) & = & {(R+a)^5-R^5 \over 32 a^2(R+a)^5} +{13 \over 512
a R} \nonumber \\
 \phi^{(2)}_2(R) & = & { 75 R^6+291 a R^5 -650 a^2 R^4 -3400 a^3 R^3
 -4800 a^4 R^2 -2925 a^5 R -669 a^6 \over 400 R^3 (R+a)^5} \nonumber \\
 &  & + {21 a\over 20 R^3} [\ln(2a)-\ln(R+a)] + {121 a \over 128 R^3} 
\label{misnerbound}
\end{eqnarray}
Notice that $\phi^{(2)}_2(R)$ falls off only as $R^{-2}$ for large
$R$. 

For completeness, we also present the solution one obtains if one
chooses the ``puncture'' boundary condition
considered by Brandt and Br\"ugmann \cite{BrBr} recently,
\begin{eqnarray}
 \phi^{(2)}_0(R) &=&
{\frac {{M}^{4}+10\,{M}^{3}R+40\,{M}^{2}{R}^{2}+80
\,M{R}^{3}+80\,{R}^{4}}{8\,M\left (2\,R+M\right )^{
5}}} \label{brbrbrbr}\\
 \phi^{(2)}_2(R) &=&{\frac {120\,{R}^{5}+768\,{R}^{4}M+1078\,{R}^{3}{M}
^{2}+658\,{R}^{2}{M}^{3}+189\,R{M}^{4}+21\,{M}^{5}}
{20\,{R}^{2}\left (2\,R+M\right )^{5}}}\\
&&-21{\frac {
\left (\ln (2\,R+M)-\ln (M)\right )M}{40\,
{R}^{3}}}\nonumber
\end{eqnarray}
but we will  only consider the solution 
(\ref{misnerbound}) in the rest of this paper. 

To obtain an initial data set appropriate for a perturbative evolution
we proceed as follows. First, in an ADM type decomposition, we choose
our shift functions $N_a=0$, so that we have,

\begin{eqnarray}
\label{ADMeqs}
ds^2 & = & g_{ab} dx^a dx^b - N^2 (dt)^2 \nonumber \\
{\partial g_{ab} \over \partial t} & = & -2NK_{ab}
\end{eqnarray}
Next, we assume that on our initial slice, 
at $t=0$, the 3-metric $g_{ab}$ and extrinsic curvature $K_{ab}$
are given by the above conformal flatness construction. Namely, we have
$g_{ab}=\phi^4 \delta_{ab}$ and $K_{ab}=\phi^{-2} \widehat{K}_{ab} $, with
$\widehat{K}_{ab}$ given by (\ref{Ksphe}), and $\phi$ given by (\ref{eqphi}).
(This, of course, ensures that our initial data satisfies the constraint
equations). We next change from the conformal spherical radial
coordinates $R$, to a ``Schwarzschild'' radial coordinates $r$, with
$R=(\sqrt{r}+\sqrt{r-2M})^2/4$, where $M = 2 a$, and choose $N=\sqrt{1-2M/r}$.
It can be checked that with these choices, for $P=0$ the extrinsic curvature
vanishes and we recover the Schwarzschild metric in the usual Schwarzschild
coordinates. This last expression for the metric in $(r,\theta,\phi)$
coordinates (or, rather, the initial data) has, therefore, the appropriate form 
for a perturbative treatment of the Regge - Wheeler type, as extended to second
order in \cite{physrep}.

The final form for the initial data that results from this
construction has thefollowing
multipolar components: to zeroth order in the momentum, we only have $\ell=0$
components; to first order in the linear momentum,only $\ell=1$ contributions;
to order ${\cal O}(P^2)$, we have $\ell=0,2$ contributions. All contributions
are even-parity and are analyzed in more detail in the following Sections.

\subsection{Multipolar decomposition of the initial data :
the $\ell=1$ contributions}

The contribution to the three-metric of zeroth order in $P$ is just
the Schwarzschild solution. The first apparently non-trivial
contribution is given at ${\cal O}(P^1)$ and corresponds as we
discussed in the previous subsection to an $\ell=1$ multipole. Let us
analyze the $\ell=1$ perturbations of a spherically symmetric
spacetime. In order to do this we use the traditional Regge--Wheeler
\cite{ReWh} decomposition. One starts with a background metric written
as,
\begin{equation}
 g^{(0)}_{\mu \nu} dx^{\mu} dx^{\nu} = -(1 -2M/r) \; dt^2 + (1 -2M
/r)^{-1} dr^2 + r^2 d\theta^2 + r^2 \sin^2 \theta d \varphi^2.
\end{equation}

  For axisymmetric perturbations, the general $\ell = 1$ even parity
terms take the form \cite{ReWh}
\begin{eqnarray}
 h^{(1)}_{tt} & = & (1 -2 M/r ) H_0(t,r) \cos \theta  \nonumber \\
 h^{(1)}_{tr} & = & H_1(t,r) \cos \theta  \nonumber \\
 h^{(1)}_{rr} & = & (1 -2 M/r )^{-1} H_2(t,r) \cos \theta  \nonumber \\
 h^{(1)}_{t \theta} & = &  - h_0(t,r) \sin \theta  \nonumber \\
 h^{(1)}_{r \theta} & = &  - h_1(t,r) \sin \theta  \nonumber \\
 h^{(1)}_{\theta \theta} & = &  r^2 K(t,r) \cos \theta  \nonumber \\
 h^{(1)}_{\phi \phi} & = &   r^2 \sin^2 \theta K(t,r) \cos \theta  
\end{eqnarray}

The first order metric perturbation coefficients are
not uniquely defined, but may be changed by ``gauge transformations''
of the form \cite{ReWh}
\begin{equation}
\widetilde{h}^{(1)}_{\mu\nu}  = 
h^{(1)}_{\mu\nu} -g_{\mu \nu},_{\rho} \xi_{(1)}^{\rho} -
g_{\mu \rho} \xi_{(1)}^{\rho},_{\nu} - g_{\rho \nu}
\xi_{(1)}^{\rho},_{\mu}
\end{equation}
where the gauge 4-vector $\xi_{(1)}^{\mu}$ is arbitrary, except for the
requirement of axisymmetry.

In particular, the $\ell = 1$ even parity coefficients transform as
\begin{eqnarray}
\label{gauge1}
\widetilde{H}_0(t,r) & = &    H_0(t,r)  +2 {\partial {\cal{M}}_0(t,r)
\over \partial t} + {2M \over r(r-2M)} {\cal{M}}_1(t,r)   \nonumber \\
 \widetilde{H}_1(t,r) & = &   H_1(t,r)  +{r-2M \over r}{\partial
 {\cal{M}}_0(t,r) \over \partial r} - {r \over r-2M} {\partial
 {\cal{M}}_1(t,r) \over \partial t}   \nonumber \\
 \widetilde{H}_2(t,r) & = &    H_2(t,r) +{2M \over r(r-2M)}
{\cal{M}}_1(t,r) -2  {\partial {\cal{M}}_1(t,r) \over \partial
r}\nonumber \\
 \widetilde{h}_0(t,r) & = &   h_0(t,r) +{r-2M \over r} {\cal{M}}_0(t,r)
 -r^2 {\partial {\cal{M}}_2(t,r) \over \partial t} \nonumber \\
 \widetilde{h}_1(t,r) & = &   h_1(t,r)  -{r \over r-2M}{\cal{M}}_1(t,r)
 -r^2  {\partial {\cal{M}}_2(t,r) \over \partial r}\nonumber \\
 \widetilde{K}(t,r) & = &   K(t,r) -{2 \over r} {\cal{M}}_1(t,r) +2
 {\cal{M}}_2(t,r)
\end{eqnarray}
where the functions ${\cal{M}}_i$ are arbitrary.

One can use this gauge freedom to go to a restricted gauge in which 
$h_0 = h_1 = K =0$. This gauge is not completely fixed. One still can
perform gauge transformations of the form,
\begin{equation}
{\cal{M}}_0 = {r^3 \over r -2M} {\partial {\cal{M}}_2  \over \partial
t}
\;\;\;,\;\;\;
{\cal{M}}_1 = r {\cal{M}}_2  \;\;\;,\;\;\;
{\cal{M}}_2 = {f(t) \over r-2M}
\label{eqxi1}
\end{equation}
where $f(t)$ is arbitrary.

An interesting result, is that in this gauge it is
straightforward to find the {\em general} solution of the linearized
Einstein equations for $\ell=1$ even parity perturbations. The result
is, 
\begin{eqnarray}
H_1 & = &  - {r \over (r-2M)^2} {d F_1 \over dt} \nonumber \\
H_0 & = &  {1 \over 3(r-2M)^2} F_1 +{r^3 \over 3M(r-2M)^2} {d^2 F_1
\over dt^2}  \nonumber \\
H_2 & = & {1 \over (r-2M)^2} F_1(t), 
\end{eqnarray} 
where $F_1(t)$ is an arbitrary function. Remarkably, one can show that
this solution is pure gauge. Choosing $f(t)=-F_1(t)/(6M)$ in
(\ref{eqxi1}) leads to vanishing gauge transformed quantities.

We therefore see that the $\ell=1$ perturbations are pure gauge. This
result was first noticed by Zerilli \cite{zerilli}. This was to be
expected in physical grounds since one could always imagine setting
coordinates boosted in such a way that the black hole would not move.
It has the further implication that we can compute the radiated energy
by computing the $\ell=2$ perturbations in a gauge in which the
$\ell=1$ perturbations vanish, and studying their evolution.

\subsection{Multipolar decomposition of the initial data :
the $\ell=2$ contributions}

The relevant $\ell=2$ perturbations are of second order in our
perturbation parameter, $P$. In principle, when one works out second
order perturbations of a given metric, the evolution equations one gets
have the general form of a linear operator (similar to the one that
acts in first order giving rise to the Zerilli equation) acting on
second order quantities, equal to a quadratic ``source'' term formed
with the first order perturbations \cite{GlNiPrPucqg}. The source term
is complicated and is delicate to handle numerically when evolving the
perturbations. A place where this was explicitly done was for instance
in the evolution of boosted black hole collisions to second order
\cite{GlNiPrPuboost2}. The calculations are lengthy and complicated.
Fortunately, in our case one can proceed in a different
way. We have just shown that there is a gauge in which the first order
perturbations (the $\ell=1$ ones) vanish.  Therefore in that gauge one
can write a second order Zerilli equation that is source-free.
Moreover, the linear portion of the second order perturbative equation
is exactly the same as the first order perturbative equation (the
Zerilli equation), and this equation can be written in terms of
quantities that are gauge invariant.  We notice, however, that
eliminating the first order $\ell=1$ terms through a first order gauge
transformation introduces second-order changes in the metric that are
not a pure second order gauge transformation, and  must be taken into
account. One must then be careful in making sure that the initial data
for the $\ell=2$ perturbations used in the Zerilli equation corresponds
precisely to that gauge. In other words, we need to carry out a first
order gauge transformation on the initial data that provides a new
initial data corresponding to a gauge where the $\ell=1$ perturbations
vanish. Since all perturbations satisfy equations that are second order
in time, this requires that the $\ell=1$ terms of the metric, and their
first time derivative vanish on the initial slice. If we consider
(\ref{gauge1}), this requires that the gauge vector components
${\cal{M}}_i$ are such that both the left hand sides in
(\ref{gauge1}), and their first time derivatives, vanish when evaluated
at $t=0$. This, in principle requires only the knowledge of the metric functions on the right hand side of (\ref{gauge1}), and their first time derivative, at the fiducial time $t=0$. It turns out, however, that we also need second order time derivatives,

 to implement the second order gauge transformation required to obtain the $\ell=2$ initial data. These second order time derivatives may be straighforwardly evaluated from the corresponding Einstein equations for the $\ell=1$ perturbations.

For the particular case in question, the $\ell=1$ initial data (and
first time derivative) is determined by (\ref{Ksphe}), and one can use
this information to construct the space-time solution of the Einstein
equations produced by the initial data as a Taylor expansion in $t$, up
to the appropriate order. One gets, using the usual Regge-Wheeler
notation \cite{ReWh} the following expansions for the $\ell=1$
components of the metric,

\begin{eqnarray}
H_2 &=& -{3\over 2} {(\sqrt{r}+\sqrt{r-2M})^2 \sqrt{r-2M}\over r^{7/2}} P t
+O(t^3)\\
K&=&{3\over 4} {(\sqrt{r}+\sqrt{r-2M})^2 \sqrt{r-2M}\over r^{7/2}} P t
+O(t^3)\\
h_1 &=& 
-{3\over 4} {(\sqrt{r}+\sqrt{r-2M})^2 \over r^{2}} P t
+O(t^3)
\end{eqnarray}
all the other components being $O(t^3)$.

The components of the gauge vector generating the first order gauge transformation that makes the initial $\ell=1$ data purely $O(t^3)$ are,
\begin{eqnarray}
\xi_t &=& {\cal{M}}_0 \cos\theta\\
\xi_r &=& {\cal{M}}_1 \cos\theta\\
\xi_\theta &=& -{\cal{M}}_2 \sin\theta\\
\xi_t &=& 0,
\end{eqnarray}
where,
\begin{eqnarray}
{\cal{M}}_0 &=&
-{\frac {P\left (-3\,\sqrt {r-2\,M}rM-5\,{r}^{3/2}M+2\,{M}^{2}\sqrt {r}
+2\,{r}^{5/2}+2\,\sqrt {r-2\,M}{r}^{2}\right )}{4\,\sqrt {r}M\left (r-2\,
M\right )}}\\
{\cal{M}}_1 &=& 
-{\frac {P\left (-6\,\sqrt {r-2\,M}rM+2\,{r}^{5/2}-8\,{r}^{3/2}M+8\,{M}^{2}
\sqrt {r}+3\,\sqrt {r-2\,M}{M}^{2}+2\,\sqrt {r-2\,M}{r}^{2}\right )t}{4\,
{r}^{5/2}M}}\\
{\cal{M}}_2 &=&
-{\frac {P\left (-3\,\sqrt {r-2\,M}rM-5\,{r}^{3/2}M+2\,{M}^{2}\sqrt {r}
+2\,{r}^{5/2}+2\,\sqrt {r-2\,M}{r}^{2}\right )t}{4\,{r}^{7/2}M}}.
\end{eqnarray}

Performing a first order gauge transformation with this generator, one
eliminates the first order $\ell=1$ component of the metric. Therefore the 
leading terms in the initial data become second order. The latter have two
contributions, both of $\ell=2$ multipolar order. One contribution simply
comes from the expansion to second order of the initial data generated via
the conformal approach. The other contribution comes from the fact that
the first order gauge transformation we just performed has second order 
pieces of the form,
\begin{eqnarray}
\widetilde{h}^{(2)}_{\mu\nu} & = &
h^{(2)}_{\mu\nu} \nonumber \\
& &  
 -\widetilde{h}^{(1)}_{\mu
\nu},_{\rho} \xi_{(1)}^{\rho} - \widetilde{h}^{(1)}_{\mu \rho}
\xi_{(1)}^{\rho},_{\nu} - \widetilde{h}^{(1)}_{\rho \nu}
\xi_{(1)}^{\rho},_{\mu} \\
& & - 
{1\over2}g_{\mu \nu},_{\sigma},_{\lambda} \xi_{(1)}^{\sigma} \xi_{(1)}^{\lambda} 
- g_{\mu \lambda},_{\sigma} \xi_{(1)}^{\sigma}
\xi_{(1)}^{\lambda},_{\nu} - g_{\lambda \nu},_{\sigma}
\xi_{(1)}^{\sigma} \xi_{(1)}^{\lambda},_{\mu} - g_{\sigma \lambda}
\xi_{(1)}^{\sigma},_{\mu} \xi_{(1)}^{\lambda},_{\nu} \nonumber
\end{eqnarray}

The contribution to the second order Zerilli function coming from the
nonlinear terms in the gauge transformation can be found after a
straightforward but tedious calculation. The relevant contributions
to the second order $\ell=2$ metric at $t=0$, are,
\begin{eqnarray}
{}^{(2)}K &=& {2 \over 3} {P^2 R^2 \over M^2 r^3} \nonumber\\
{}^{(2)}H_2 &=&  {2 \over 3} {P^2R^2 (R-M)^2 \over M^2 r^3 (r-2M)}
\nonumber\\
{}^{(2)}h_1 &=& {1 \over 3} {P^2 R^{5/2} (R-M) \over M^2 r^{3/2}
(r-2M)} \nonumber\\
{}^{(2)}G &=& {1 \over 3} {P^2 R^2 \over M^2 r^3}
\end{eqnarray}
where $R = (\sqrt{r}+\sqrt{r-2M})^2/4$.


Another contribution to this initial data comes from the second order
corrections to the conformal factor, calculated above.  These
contributions to the second order $\ell=2$ metric all vanish, except
for,
\begin{eqnarray}
{}^{(2)}H_2 & = & 4 P^2 \left(  1 + {M \over 2R}
 \right)^3 \phi^{(2)}_2(R) {R^2 \over r^2} \nonumber \\
{}^{(2)}K & = & {}^{(2)}H_2, 
\end{eqnarray}
and all first order time derivatives vanish.

The Zerilli function in an arbitrary gauge is given by \cite{moncrief},

\begin{eqnarray}
  \psi(t,r) & = & \frac{r(r-2M)}{3(2 r +3M)}\left[
{\ }^{(2)}{H}_2
-r\frac{\partial {\ }^{(2)}{K}}{\partial r}
-\frac{r-3M}{r-2M}{\ }^{(2)}{K}
\right] \nonumber \\
& & +{r^2 \over 2 r +3M} \left[{\ }^{(2)}{K}
+(r-2M)
\left(
\frac{\partial {\ }^{(2)}{G}}{\partial r}
-{2 \over r^2}{\ }^{(2)}{h}_1
\right) \right] .
\end{eqnarray}

Substituting the above components yields the following initial data for the 
Zerilli function,
\begin{equation}
\psi(t=0,r)  =  
P^2\left[ {(3M+8R)R^2 \over 6M^2r(2r+3M)} + {4 r (2M+3R)
\over 3 (2r+3M)} \phi^{(2)}_2(R)- {(4r-8M) R^{1/2} r^{3/2} \over 3
(2r+3M)} \left({\partial \phi^{(2)}_2(R) \over \partial r}\right)
\right]
\end{equation}

\begin{equation}
\left.{\partial \psi(t,r) \over \partial t}\right|_{t=0} =0.
\end{equation}

For the conformal factor we computed in (\ref{misnerbound}), this results in,
\begin{eqnarray}
\psi(t=0,r)  &=& 
{P^2 R^6 \over M (M+2R)^4 (M^2+10MR+4R^2)} \left(
{128 \over 3} +56 {M \over R} +{6113 \over 25} {M^2 \over R^2} + 
{32607 \over 100} {M^3 \over R^3} \right. \nonumber \\
& & \left. +{3882 \over 25} {M^4 \over R^4} -{235 \over 16} {M^5 \over R^5} 
- {691 \over 16} {M^6 \over R^6} - {26679 \over 1600} {M^7 \over R^7} -
{2047 \over 800} {M^8 \over R^8} -{2327 \over 19200} {M^9 \over R^9} \right)
\nonumber \\
& & +{7 P^2 M (M+12R)(M+2R)^2(\ln(2M)-\ln(M+2R)) \over 40 R^3 (M^2+10MR+4R^2)}.
\end{eqnarray}

\subsection{Physical validity of the perturbative treatment}

Inspecting the initial data for the Zerilli function one finds that it
behaves in a non-conventional way. Here is where we note significant
differences with the case of single spinning holes \cite{GlNiPrPuboyo}.

The first unusual thing one notices is that the initial data goes
asymptotically to a constant value for $R\rightarrow\infty$. This is
different from the data for the ``close limit'' of two black holes
(momentarily stationary or boosted) where the Zerilli function goes to
zero at infinity. The root of this problem can be traced down to the
falloff conditions of the extrinsic curvature. In all other cases in
which the ``close limit'' approximation has been applied, the
extrinsic curvature falls off as $1/r^3$ at infinity. For the case in
this paper, it decreases as $1/r^2$ (otherwise the momentum would
vanish). This, in particular, leads to falloff conditions in the gauge
vectors we use to eliminate the $\ell=1$ even pieces of the extrinsic
curvature. In turn, the falloff condition of the gauge vectors
influences, via the quadratic terms in the second order gauge
transformation, the behavior of the Zerilli function we obtain at the
end of the process.

Does the appearance of a Zerilli function that does not vanish at
infinity indicate something problematic in itself? We do not seem to
see any difficulty in evolving the problem in this case. Our evolution
code evolves a ``slab'' region consistent of the initial data and its
domain of dependence, therefore we do not need to specify any boundary
conditions. If one observes the behavior of the Zerilli function as a
function of time for a fixed (finite) value of the radius, it starts
having a constant value followed by a quasinormal ringing and a power
law tail decrease towards zero. That is, the constant behavior at
infinity translates itself in a certain behavior at the beginning of
the waveform, it does not leave any visible effect after the ringdown
and power law tail behavior. The radiated energy {\em and all
observable quantities at infinity} (even to second order in
perturbation theory, see section III.B of \cite{physrep}) are not
determined by the Zerilli function itself but by its time derivative,
for which the initial data indeed goes to zero for large values of the
radius, and therefore no problem in the evaluation of physical
quantities is present. This might be at first surprising, but in
reality the Zerilli function deos noe play any physical role, it is
its time derivative what does, and the latter does not have any
unusual asymptotic behavior. To put it in different terms: one could
carry out perturbation theory and work out all relevant physics
entirely in terms of the time derivative
of the Zerilli function (which also satisfies the Zerilli equation)
and there one would not see any unusual asymptotic behavior.

Another aspect that could cause concern is the nature of the gauge
transformation considered. The gauge transformation is well behaved in
any finite value of $r_*$ (that is, any point of the exterior of the
black hole excluding the horizon and spatial infinity). Since in order
to compute the radiated energies and waveforms we will never need
information from either the horizon nor infinity, the gauge
transformation is well defined in all relevant points for our
calculation.

\section{Results and conclusions}

We have numerically evolved the Zerilli equation with the initial data
presented above and computed the energy and waveforms for the
``relaxation'' of a single boosted Bowen--York hole to a boosted
Schwarzschild black hole. The results for the energy can be summarized
by a simple formula which, to leading order in $P$ reads,
\begin{equation}
\left({E_{\rm radiated} \over M_{ADM}}\right) = 4.1 \times 10^{-2} 
\left({P\over M_{ADM}}\right)^{4}
\end{equation}
where the prefactor was computed numerically
\footnote{The ADM mass of the slice depends on the value of the
momentum. For these calculations we are using the zeroth order
approximation to the ADM mass, which is constant. It is known from
full numerical calculations that for $P/M_{ADM}<0.6$ the discrepancy
between the zeroth order approximation and the full value is less than
a $1\%$ (see figure 1 of reference \cite{CoYo}). This can also be seen
from the perturbative calculation of the mass, which one can obtain
from the conformal factor we computed, and yields $M_{ADM} = M+ {P^2
\over 8M} - {2287 P^4 \over 240000 M^3}$, for instance, for the
``puncture'' case corresponding to the conformal factor of equation
(\ref{brbrbrbr}).}.

\begin{figure}
\centerline{\hbox{\psfig{figure=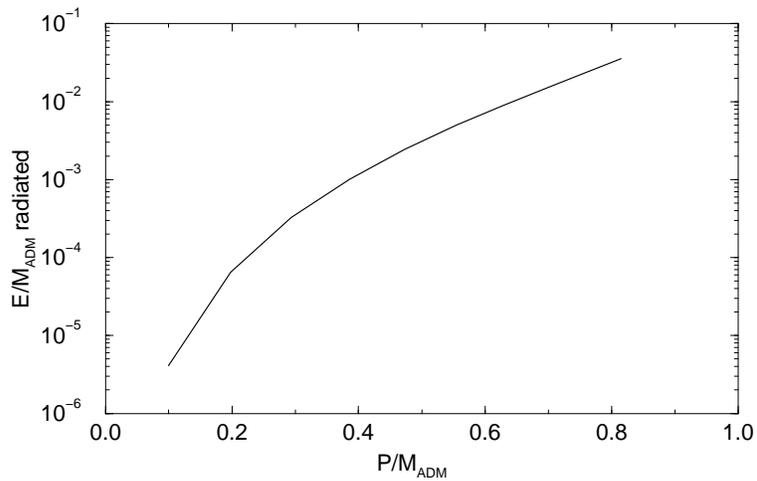,height=2.5in}}}
\caption{The energy radiated in the ``relaxation of a single, boosted
Bowen--York black hole to a Schwarzschild black hole, as a
function of the momentum. At about $P/M_{ADM}\sim 0.4$ the radiation
is equivalent to the total energy radiated in a the ``close limit''
collision of two boosted black holes. } 
\label{fig1}
\end{figure}

\begin{figure}
\centerline{\hbox{\psfig{figure=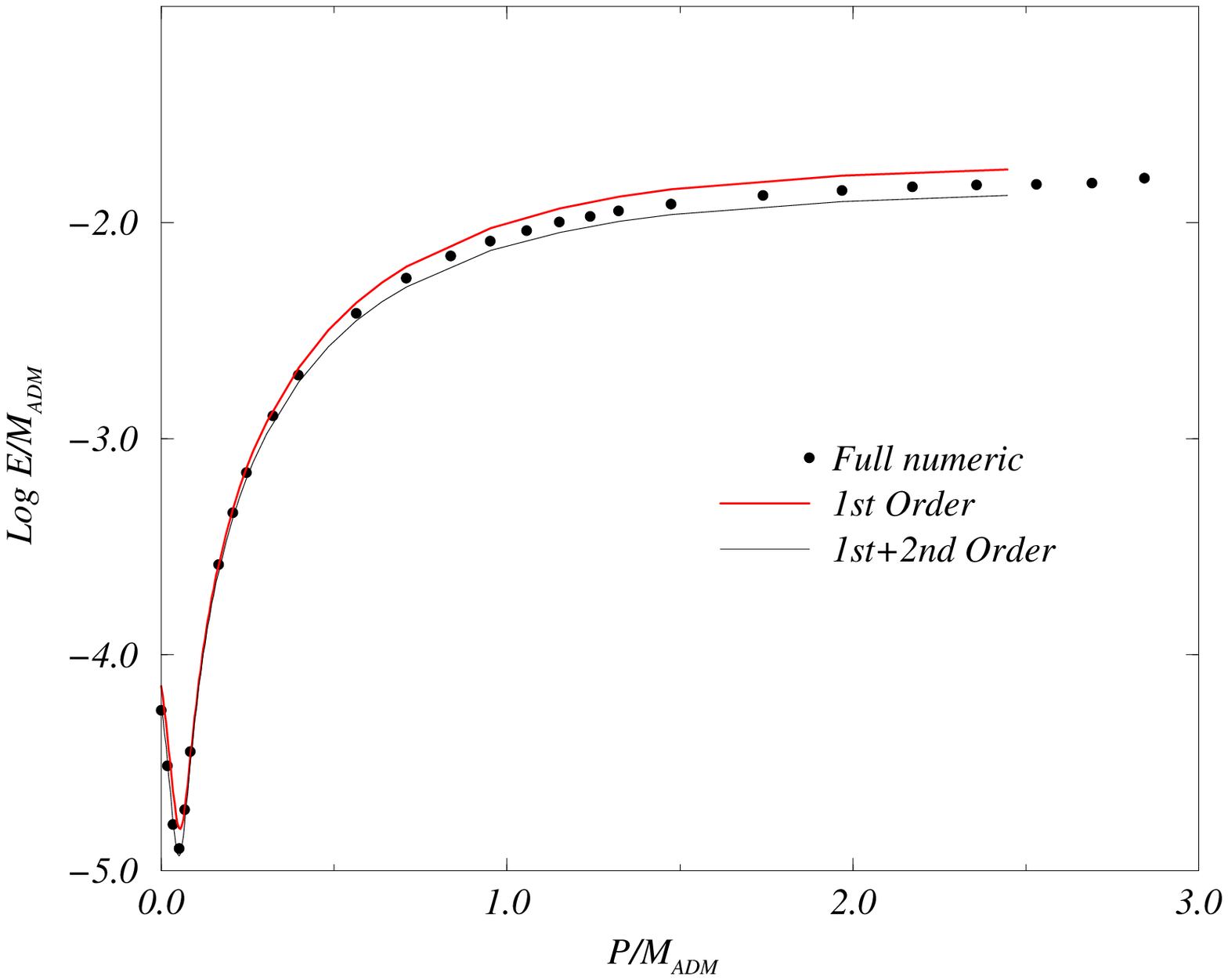,height=2.5in}}}
\caption{The energy radiated in the head on collision of two boosted
black holes as calculated in reference 10. The collision is for two
black holes separated by a Misner parameter $\mu_0=1.5$. Depicted are
the results of first and second order perturbation theory and the full
numerical results of the NCSA-Potsdam-WashU group. We see that the
collisions never radiate more than $1\%$ of the ADM mass in
gravitational waves.}\label{fig2}
\end{figure}

Figure \ref{fig1} shows
the radiated energy as a function of the momentum. We see that for
values of the momentum close to $P/M_{ADM}\sim 0.4$ the total radiated
energy by the ``relaxation'' of a single boosted Bowen--York black
hole to a Schwarzschild black hole is similar to the total radiated
energy in a close limit collision of Bowen--York black holes
\cite{GlNiPrPuboost2} and figure \ref{fig2}. One could therefore be
tempted to say that for values of $P/M_{ADM} > 0.4$ one should stop
using these families of initial data. A puzzling element is that we
have already evolved these families of initial data in collision
situations and compared with actual non-linear integrations of the
Einstein equations \cite{boost1,GlNiPrPuboost2} and {\em we know} that
these families of initial data radiate less in collisions than the
values we are predicting here for each individual hole, at least for
close separations. The results for these collisions are recollected in
figure \ref{fig2}.

What is going on? One has to keep these results in perspective, since
it is easy to get carried away and believe that perturbation theory
should work way beyond where it was meant to do so. In the calculation
of the present paper we find that the radiated energy goes as
$P^4$. This is good, since our perturbative parameter is $P$ and
therefore this means that the corrections are small. In the case of
colliding black holes the radiated energies contain terms that go as
(see reference \cite{GlNiPrPuboost2}) $L^4$, $P^2 L^2$ and $P L^3$,
where $L$ is the separation of the black hole centers in the
conformally related flat space. If one simply increases the value of
$P$ keeping $L$ constant, it is obvious that the contribution we
consider in this paper will quickly dominate. However, one is clearly
pushing things beyond the realm in which these calculations were meant
to be reliable. One is essentially forcing {\em a higher order term in
perturbation theory} to a regime in which it is larger than the lower
order terms!

Therefore the conclusion of this paper has to be read in the following
way: the terms coming from the ``relaxation'' of a conformally flat
hole to a usual boosted slice of Schwarzschild are higher order in
perturbation theory than the ones one obtains in a collision. Because
of this fact, they grow fast with the perturbative parameter and
perturbation theory breaks down early for the estimation of the
involved energy in the calculations of the current paper. The
breakdown occurs earlier for a single hole than for a collision of
holes (at least computed in the center-of-momentum frame).

Connected with the latter point is another interesting insight gained
from the analysis of this paper. It has to do with the choice of frame
used to describe collisions of black holes in perturbation theory. The
situation is illustrated in figure \ref{fig3}. One would expect these
two collisions to be physically equivalent. However, if one considers
conformal black holes and expands in perturbation theory, the
collision at the bottom will contain terms that behave exactly like
those we discuss in the current paper, whereas the top collision does
not. If we power count, for the bottom collision, the extrinsic
curvature has leading terms that behave like $P$ and are $\ell=1$ and
the conformal factor has terms that behave like $L^2$ plus terms that
behave like $P^2$ at leading order. The energy, being quadratic in the
fields, will generically contain terms that behave like $L^4$, $P^2
L^2$ and $P^4$, the latter being the terms we encountered in this
paper. However, the top collision contains terms $L^4$, $P L^3$ and
$P^2 L^2$, as discussed in reference \cite{GlNiPrPuboost2} (the
extrinsic curvature goes as $PL$ and the conformal factor as $L^2$ at
leading order). Therefore if one were to consider ``close'' black
holes and were to consider the radiated energy as a function of $P$,
as we have done above, one would encounter that the perturbative
predictions ---at the order considered--- will differ significantly as
soon as the $P^4$ terms start to grow. The moral from this paper
insofar as the choice of the origin is: perturbation theory breaks
down quickly as a function of momentum for situations with net linear
momentum, it is best to analyze collisions set up in the
center-of-momentum frame.

As expected, given the nature of the Zerilli equation, the waveforms
that one obtains from the ``relaxation'' just behave like quasinormal
ringing.  Figure  \ref{fig4}
shows the waveform of the decay. The form is that of a
typical ringdown, and it therefore takes an amount of time of the
order of the light-crossing time of the black hole size to
decay.

Summarizing, as in the case of spinning Bowen--York black holes, one
has extra radiation present in the initial data, that grows with the
value of the momentum. In evolutions of binary black hole collisions,
one can either wait long enough for this energy to be ``flushed away''
from the system, or restrict oneself to values of the momentum that
are small enough that the extra energy is small respect to the total
energy produced in the collision. The latter is the only option in the
case of ``close limit'' collisions.  Another conclusion is that
perturbation theory breaks down quickly as a function of the momentum
of the holes for single boosted holes (and collisions of black holes
not computed in the center-of-momentum frame) and therefore cannot be
used to reliably estimate the ``energy content'' of each hole in a
regime that might be of interest for the momenta and energies relevant
for black hole collisions.

\section{Acknowledgments}

We wish to thank a anonymous referees for comments on an earlier
version of this manuscript.  This work was supported in part by grants
of the National Science Foundation of the US INT-9512894, PHY-9423950,
PHY-9800973, PHY-9407194, by funds of the University of C\'ordoba, the
Pennsylvania State University and the Eberly Family Research Fund at
Penn State. We also acknowledge support of CONICET and CONICOR
(Argentina). JP also acknowledges support from the Alfred P. Sloan and
John S. Guggenheim Foundations.  JP was a visitor at ITP Santa Barbara
during the part of the completion of this work. The authors wish to
thank Richard Price and the University of Utah for hospitality.
This work was supported in part by funds of the Horace Hearne Jr. 
Institute for Theoretical Physics.



\begin{figure}
\centerline{\hbox{\psfig{figure=fig2.eps,height=1.5in}}}
\caption{Two black hole collisions that should be physically
equivalent, but that are significantly different from the point of
view of the perturbative treatment involved in the ``close limit
approximation''.}\label{fig3}
\end{figure}

\begin{figure}
\centerline{\hbox{\psfig{figure=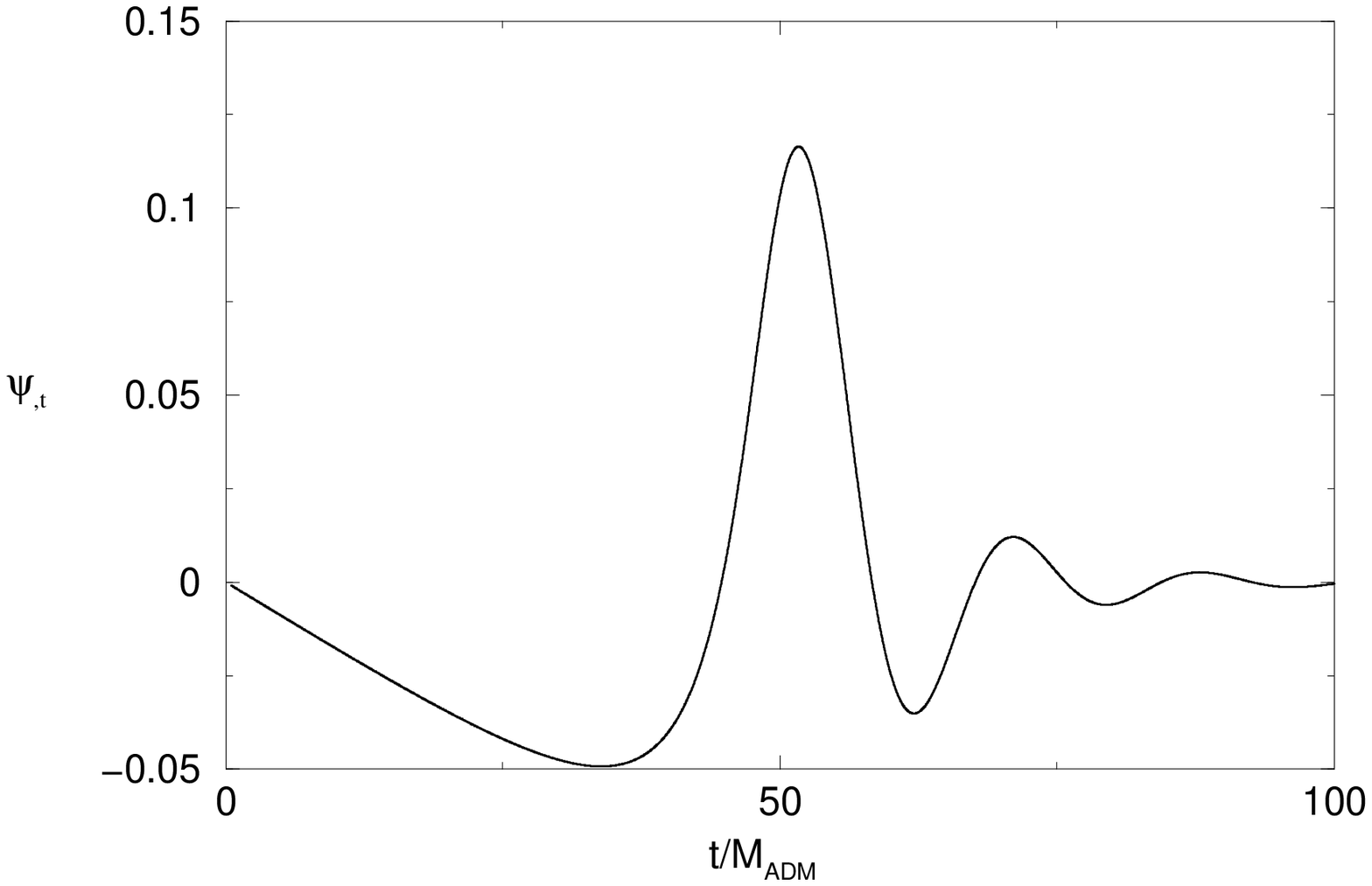,height=2.5in}}}
\caption{The time derivative of the Zerilli function as a function of
time. The square of this quantity is proportional to the energy flux,
and therefore characterizes the ``waveform'' of the gravitational
radiation from the decay of a Bowen--York boosted black hole into
Schwarzschild.}\label{fig4}
\end{figure}
\end{document}